\documentclass[11pt]{cernrep}
\usepackage{graphicx}

\newcommand{\be}{\begin{eqnarray}}
\newcommand{\ee}{\end{eqnarray}}

\begin{document}

\title{THE SIGN PROBLEM IS THE SOLUTION}

\author{J.C. Osborn$^1$, K. Splittorff$^2$ and J.J.M. Verbaarschot$^3$}

\institute{$^1$ Physics Department, Boston University,
Boston, MA 02215, USA \\ $^2$ The Niels Bohr Institute, Blegdamsvej 17, DK-2100,
Copenhagen {\O}, Denmark \\ $^3$  Department of Physics and Astronomy, SUNY, Stony Brook,
 New York 11794, USA}

%

\maketitle

\begin{abstract}
The unquenched spectral density of the Dirac operator at $\mu\neq0$ is
complex and 
has oscillations with a period inversely proportional to the volume and an
amplitude that grows exponentially with the volume. Here we show how
the oscillations lead to the discontinuity of the chiral condensate.     
\end{abstract}

\section{Lessgo !}

The sign problem in QCD at non-zero baryon chemical
  potential, $\mu$, has been an obstacle for more than two decades. 
New innovations have given exiting results \cite{innovative} even though 
the core of the sign problem in QCD remains.   
Here we address a closely related and equally long-standing problem 
(see e.g.~\cite{BCmu}):
\vspace{2mm} 

\noindent
{\sl How does chiral symmetry breaking at $\mu \ne 0$ manifest itself in the
  spectrum of the Dirac operator?} 
\vspace{2mm}

\noindent
At $\mu=0$ the Dirac spectrum  is located on the 
imaginary axis and the discontinuity of the chiral condensate at zero quark
mass, $m$, is proportional to the eigenvalue density at the origin (the Banks
Casher 
relation). While this is quite intuitive the situation for
$\mu\neq0$ is quite puzzling. The chemical potential 
sends the eigenvalues, $z_k$, of $D+\mu\gamma_0$ off into the complex plane 
while the discontinuity of the chiral condensate remains. The
solution \cite{OSV} of this (Silver Blaze \cite{cohen}) puzzle is in a 
sense the sign problem: Due to the sign problem the eigenvalue
density becomes a complex
and strongly 
oscillating function and the oscillations lead to the discontinuity of the
chiral condensate.  
To show this we start from the exact solution \cite{O} of 
the eigenvalue density for $\mu \ll \Lambda_{\rm QCD}$ and 
$m_\pi^2 \ll 1/\sqrt V$ ($\langle\ldots\rangle$ is
the quenched average)   
\be \label{rhoNf-def}
 \rho_{N_f}(x,y,m;\mu) =
\frac{
 \langle \sum_k \delta^2(x+iy - z_k)\,
 {\det}^{N_f}( D+\mu\gamma_0+m ) \rangle
}{\langle {\det}^{N_f}( D+\mu\gamma_0+m ) \rangle}
\ee
and compute the chiral condensate 
$\Sigma=\lim_{m\to0,V\to\infty}\Sigma_{N_f}(m)$ using 
\be\label{Sigma-from-rho}
\Sigma_{N_f}(m) = 
\frac 1{V} \int dx\,dy \, \frac{\rho_{N_f}(x,y,m;\mu)}{x+iy +m}.
\ee
Given the nature of the sign problem it is perhaps not surprising that the 
oscillations of the eigenvalue 
density (\ref{rhoNf-def}) have a period inversely
proportional to the volume and an amplitude growing exponentially large with
the volume \cite{AOSV}.  
The results \cite{SplitVerb2,O,AOSV} for the eigenvalue density 
are ideally suited to resolve these oscillations since they describe 
eigenvalues $|z|\sim {\cal O}(1/(\Sigma V))$. In \cite{O} the eigenvalue 
density was derived using non hermitian random matrix theory \cite{Gernot} 
while in \cite{AOSV} it was derived from the chiral Lagrangian 
using the replica method \cite{KandSV}.\\
\noindent
On the next pages we explain the original \cite{OSV} direct computation of 
$\Sigma_{N_f}(m)$ through several figures.

\begin{figure}[!ht]
\begin{center}

\begin{picture}(30,2.0)  
  \put(15,-160){\bf\large $x\Sigma V$}
  \put(190,-24.0){\bf\large $y\Sigma V$}
  \put(-200,-1.7){\bf \LARGE
$\frac{\rho_{2n=2}(x,y,m;\mu)}{\Sigma^2V^2}$}
\end{picture}

\includegraphics[width=12cm]{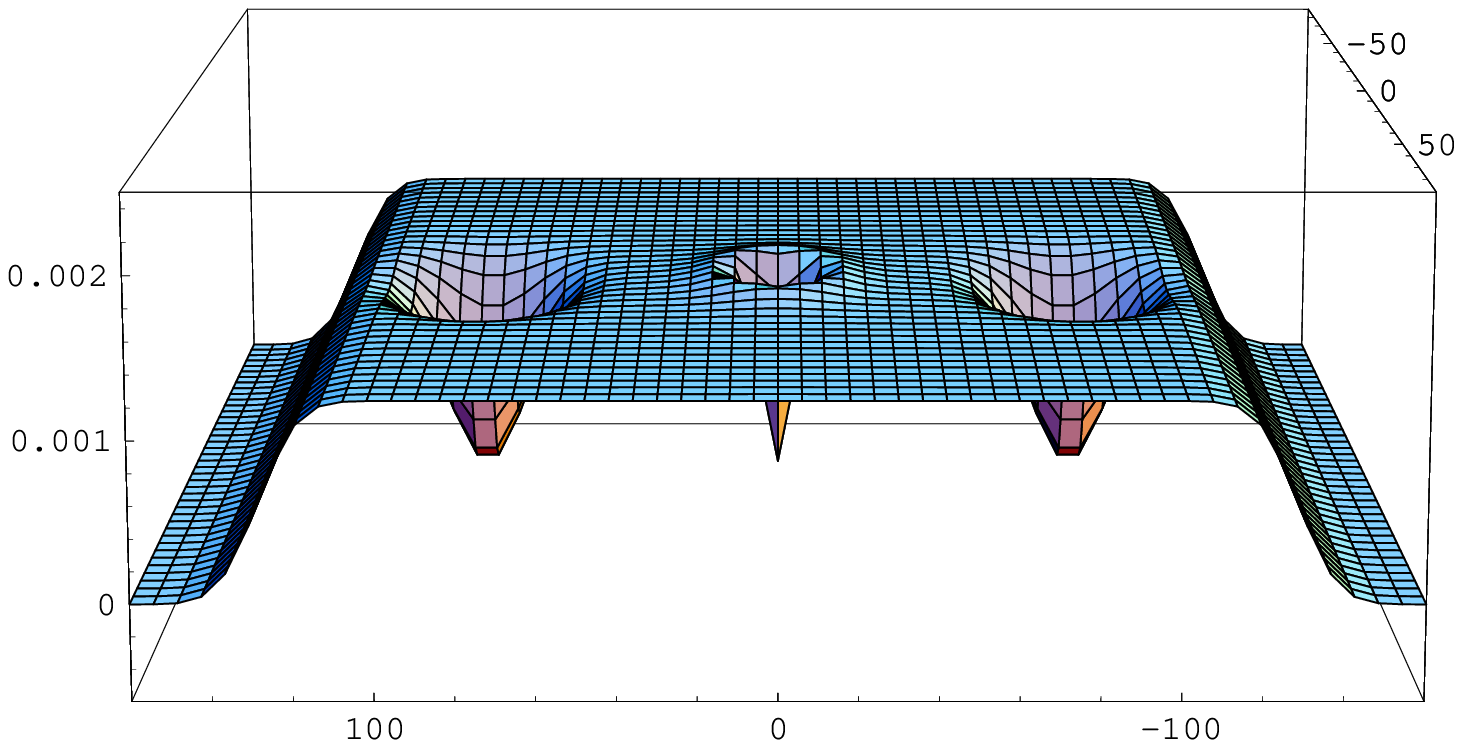}
\vfill
\hspace{-8mm}\includegraphics[width=12.8cm]{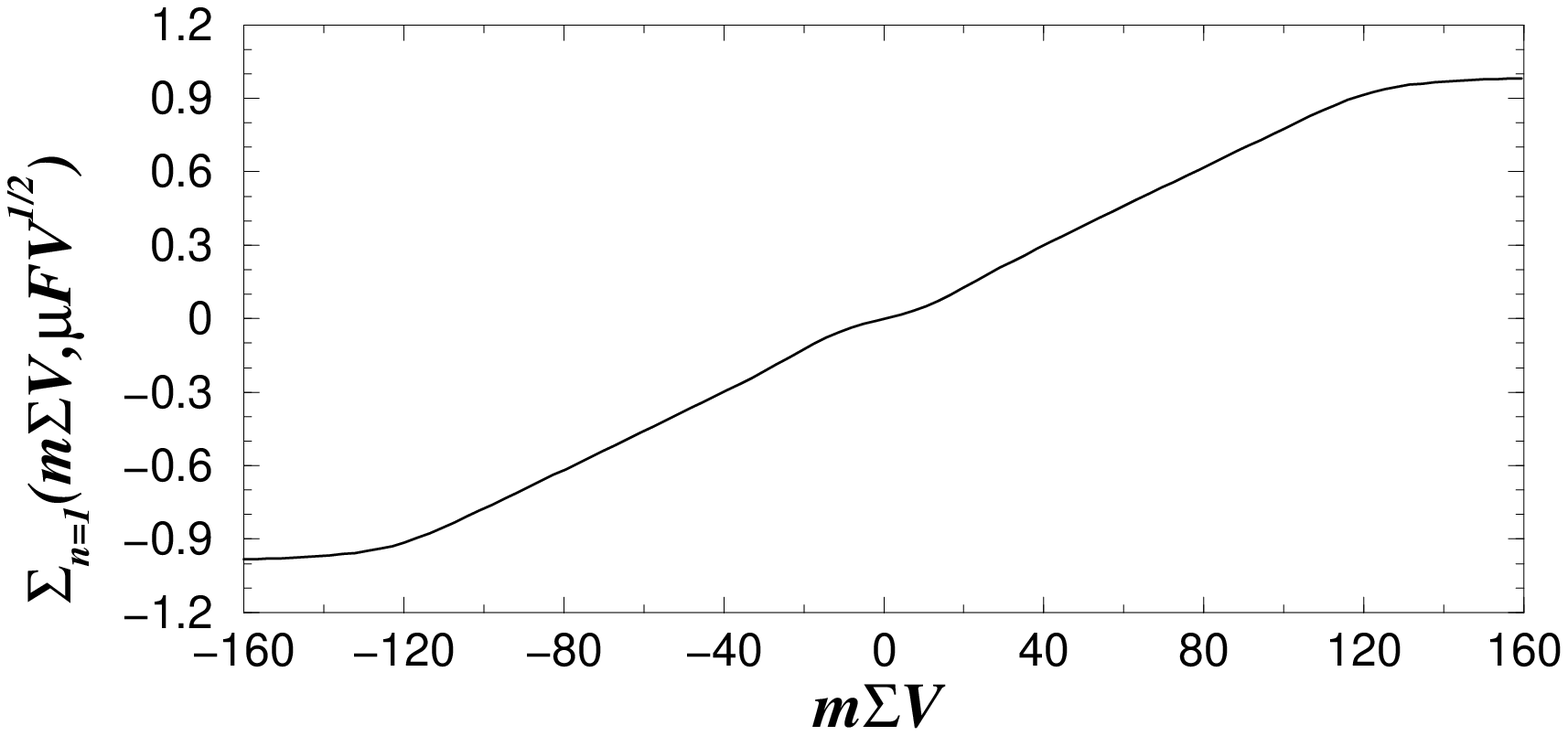}
\end{center}
\vskip -1cm
\caption{\label{fig1}{\bf Top:} The eigenvalue density in the complex
  eigenvalue plane with two phase quenched
  flavors  for $\mu F \sqrt{V}=8$ and $m\Sigma V=80$. {\bf Bottom:} The
  corresponding chiral condensate as a function of $m\Sigma V$, again for $\mu F \sqrt{V}=8$.}
\vskip -0.3cm
\end{figure}

\begin{figure}[!ht]
\begin{center}

\begin{picture}(30,2.0)  
  \put(15,-144){\bf\large $x\Sigma V$}
  \put(200,4.0){\bf\large $y\Sigma V$}
  \put(-225,10){\bf \LARGE
$\frac{Re[\rho_{N_f=1}(x,y,m;\mu)]}{\Sigma^2V^2}$}
  \put(15,-340){\bf\large $x\Sigma V$}
  \put(200,-194.0){\bf\large $y\Sigma V$}
  \put(-225,-190.7){\bf \LARGE
$\frac{\rho_{Q}(x,y,m;\mu)}{\Sigma^2V^2}$}
  \put(15,-540){\bf\large $x\Sigma V$}
  \put(200,-390.0){\bf\large $y\Sigma V$}
  \put(-225,-381.7){\bf \LARGE
$\frac{Re[\rho_{U}(x,y,m;\mu)]}{\Sigma^2V^2}$}
\end{picture}

\vspace{-1cm}\includegraphics[width=13cm]{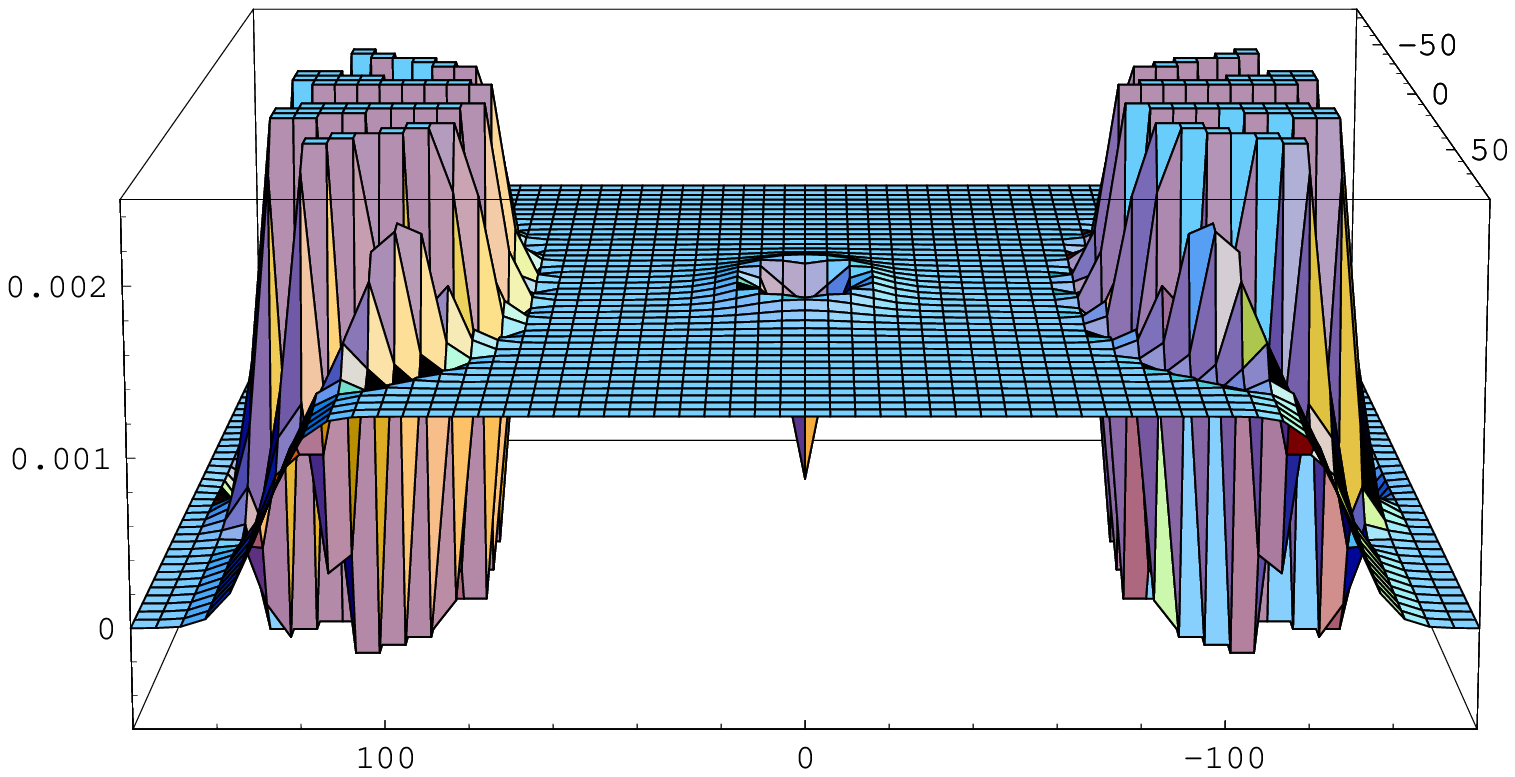}
\vfill
\includegraphics[width=13cm]{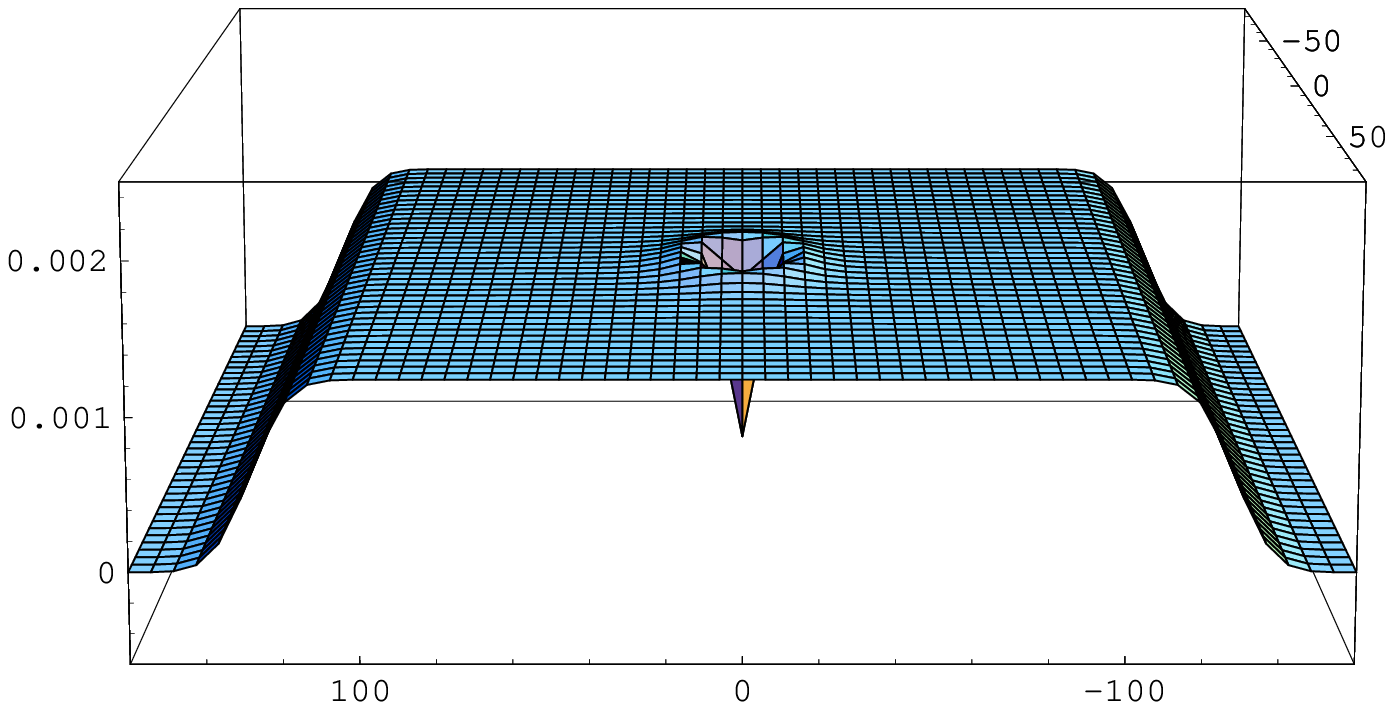}
\vfill
\includegraphics[width=13cm]{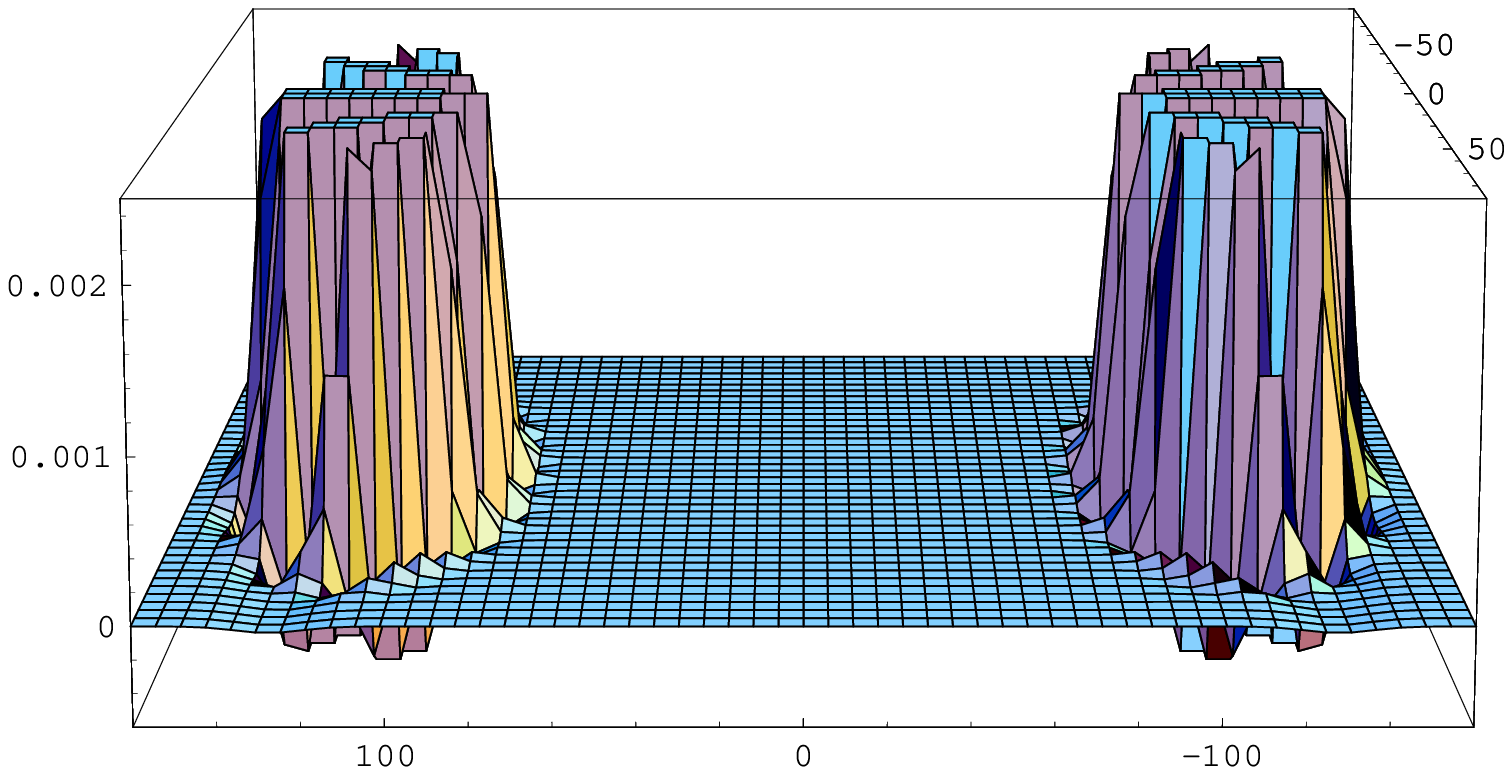}
\vskip -0.3cm
\caption{\label{fig2} The eigenvalue density of the full QCD Dirac 
operator for one flavor (real part only) 
and  $\mu F \sqrt{V}=8$ and $m\Sigma V=80$. The bottom figure shows
the difference between the full (top) and quenched (middle)  eigenvalue 
density.
Notice the similarity between quenched  and the phase
quenched spectral density shown in figure \ref{fig1}.
For $y=0$ the oscillations  start $|x|=m$.}
\vskip -0.5cm
\end{center}
\end{figure}

\begin{figure}[!ht]
\begin{center}
\vspace{-1cm}\includegraphics[width=12cm]{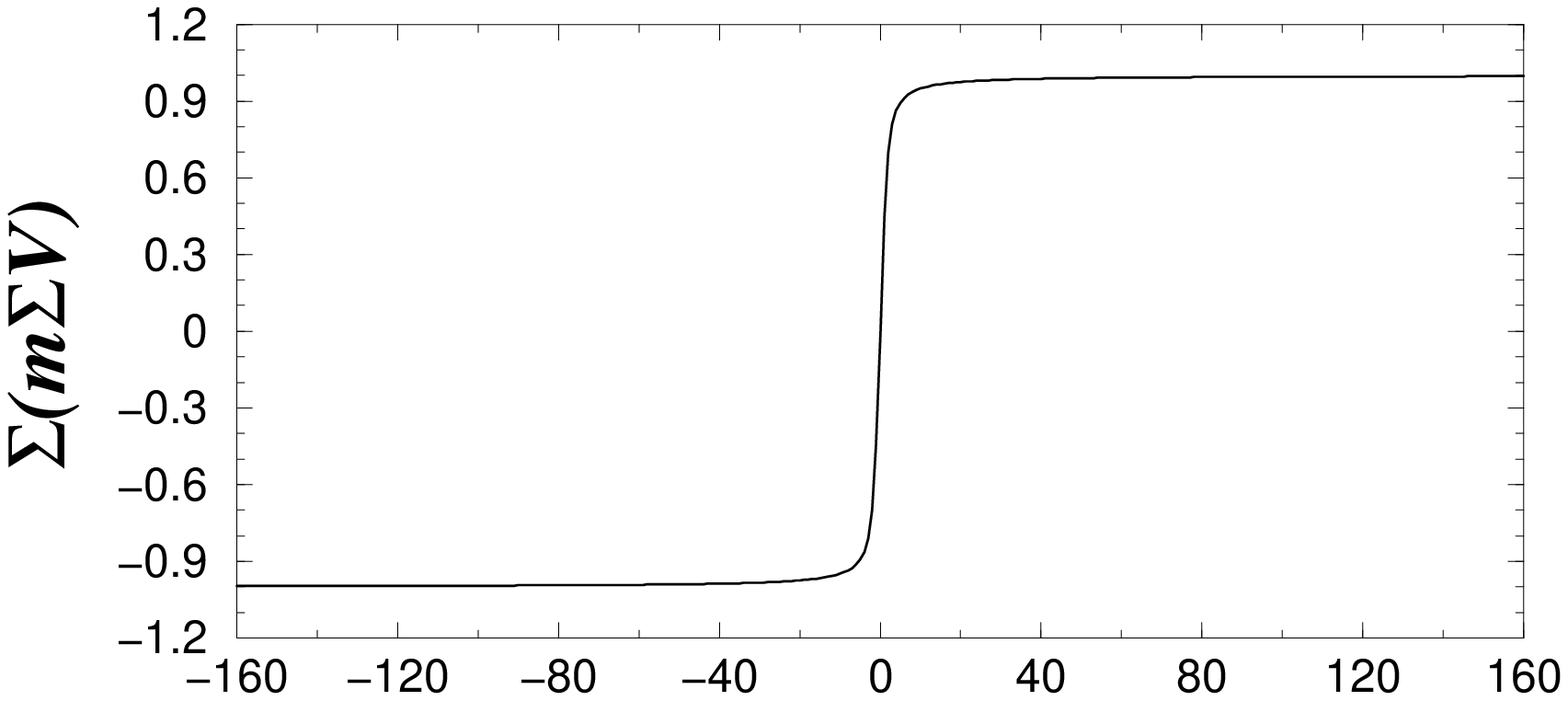}
\vfill
\vspace{-5mm}
\includegraphics[width=12cm]{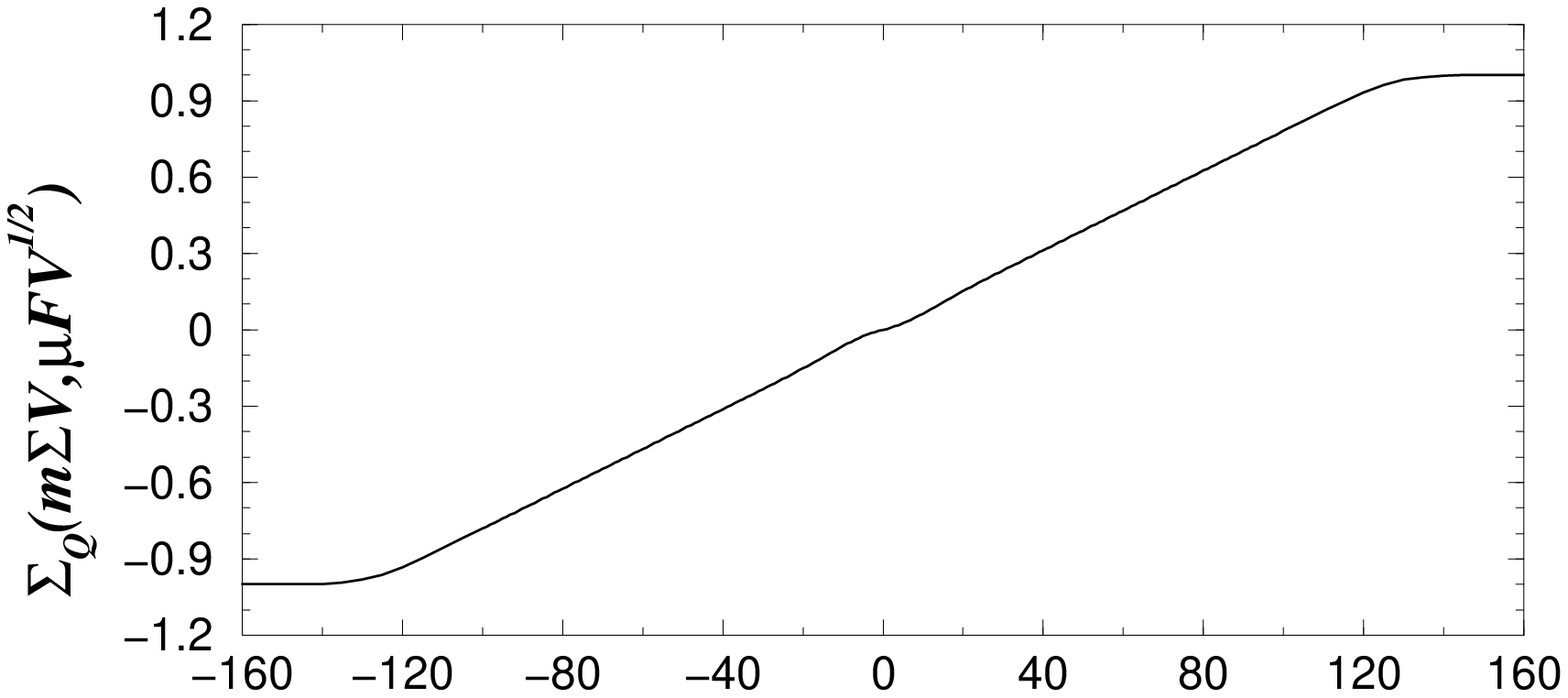}
\vfill
\vspace{-5mm}
\includegraphics[width=12cm]{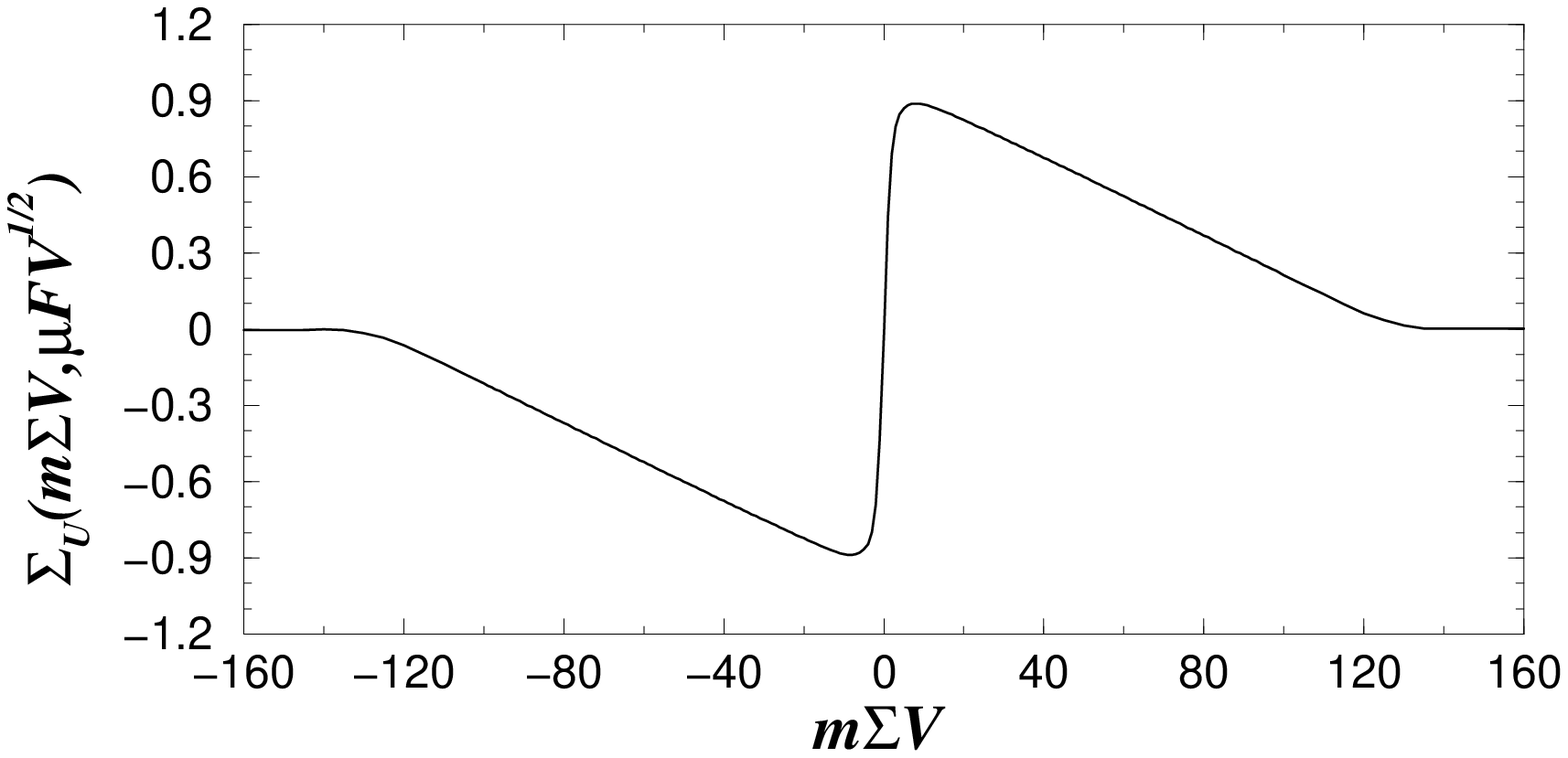}
\vskip -0.3cm
\caption{\label{fig3} The chiral condensate as a function of quark
  mass obtained from the three densities in figure \ref{fig2} using
  (\ref{Sigma-from-rho}). \newline
{\bf Top:} The chiral condensate in full QCD with the discontinuity at $m=0$
(independent of $\mu$) is the sum of the two below.
\newline
{\bf Middle:} The quenched chiral condensate (dependent on $\mu$)
\cite{OSV2}. No discontinuity at $m=0$.   \newline
{\bf Bottom:} The contribution from the oscillating part of the eigenvalue
density (dependent on $\mu$).}
\vskip -0.3cm
\end{center}
\end{figure}

\section{The phase quenched way}

Before going to the real problem we take a short aside and look at the way
chiral symmetry breaking affects the spectral density of the Dirac operator
in phase quenched QCD (tantamount to QCD at 
non-zero isospin chemical potential). This case serves to show how the perhaps more familiar 
mean field results are a special limit of the exact results. 
The real and positive eigenvalue density 
\be \label{rhoPhaseQ-def}
 \rho_{2n}(x,y,m;\mu) =
\frac{
 \langle \sum_k \delta^2(x+iy - z_k)\,
 |{\det}( D+\mu\gamma_0+m )|^{2n} \rangle
}{\langle |{\det}( D+\mu\gamma_0+m )|^{2n} \rangle}
\ee
is plotted in the top panel of figure \ref{fig1} (the explicit expression is
given in (76) of \cite{AOSV}). As expected the eigenvalues have spread out
away from the imaginary axis. 
Notice the absence of eigenvalues at $z=\pm m$ and at $z=0$. Using
(\ref{Sigma-from-rho}) the corresponding chiral condensate follows, see
the lower panel of figure \ref{fig1}.
The drop of the chiral condensate as the quark mass enters the
eigenvalue distribution is easy to understand by an electrostatic analogy and
is consistent with lattice measurements \cite{KogutSinclair}. 
Furthermore, taking $m\Sigma V\gg1$ and $\mu^2F^2V\gg1$ the mean field result 
($\Sigma^{\rm MF}_{2n}=m\Sigma^2/(2\mu^2F^2)$ for $m_\pi/(2\mu)<1$ and
$\Sigma$ otherwise)  
is reproduced by the exact result. ($F$ is the pion decay constant.)

\section{The unquenched way}

 In full QCD the eigenvalues of the Dirac
operator for a given gauge configuration are indistinguishable from
what we would have obtained in the phase quenched case. However, due to 
the
phase of the fermion determinant, the average spectral density is entirely different.
To show this we have plotted the unquenched eigenvalue 
density in the top panel of figure \ref{fig2}. Its explicit expression 
is given in (73) of \cite{AOSV}. Here it is sufficient to note that this
expression naturally separates into the quenched spectral 
density and a remainder
\be\label{refrhoNf}
\rho_{N_f}(x,y,m;\mu) =\rho_Q(x,y;\mu)+ \rho_{U}(x,y,m;\mu).
\ee
These two parts are shown in the two lower panels of figure \ref{fig2}. The
quenched eigenvalue density is real and positive and behaves as the 
phase quenched eigenvalue
density (except for the dip at the quark mass). The unquenched part contains  
the complex oscillations of which we have only displayed the real part. 
For the values of the parameters $m\Sigma V$ and $\mu^2F^2 V$
given in figure \ref{fig2},  the maximum value of the amplitude of the
oscillations is 
400 times larger than the plateau of the quenched spectral
density. For better illustration the oscillation have been clipped. 
Inserting the density (\ref{refrhoNf}) into
(\ref{Sigma-from-rho})  leads to two terms  
\be
\Sigma_{N_f}(m) = \Sigma_{Q}(m,\mu)+\Sigma_{U}(m,\mu).
\ee
The quenched part, $\Sigma_Q$, drops to zero as $m$ comes inside the
support the eigenvalue density. However, the unquenched oscillating part
exactly makes up for this and leaves the full chiral condensate
$\Sigma_{N_f}(m)$ independent of $\mu$ and therefore equal to the result for
$\mu =0$ \cite{LS} (see figure \ref{fig3}).  
While $\Sigma_Q$ is built up by the eigenvalue density inside the quark
mass (the contributions from $|x|>m$ cancel each other), the 
unquenched part $\Sigma_U$ is built up by the 
density $\rho_U$ outside the quark mass (not easy to understand by an
electrostatic analogy). This is possible since the oscillations
have a period of order $1/V$ and an amplitude that grows
exponentially with $V$ \cite{OSV}. Moreover, the contribution from the
unquenched part comes from the entire oscillating region and not  from
the boundary of the support of $\rho_U$.

\section{Conclusions} 

While the discontinuity of the chiral
condensate is usually associated with a dense spectrum of Dirac
eigenvalues on the imaginary axis near zero, the sign problem allows for 
an alternative mechanism: The discontinuity of the chiral condensate arises 
from strong oscillations of the eigenvalue density. Having a
period of order $1/V$ and an amplitude that grows exponentially large with
$V$ the oscillations directly reflect the sign problem. 
The oscillations 
are present as soon as the quark mass hits the support of the
eigenvalue density, i.e.~when the phase quenched theory enters the BEC
phase. A lattice simulation in this region therefore must be able to deal with 
the oscillations in order to study the correct mechanism for spontaneous 
chiral symmetry breaking in QCD \cite{sign}.

\section*{Acknowledgments}
It is a pleasure to thank the organizers of XQCD for hosting a friendly and inspiring
conference. JV acknowledges support by US DOE grant DE-FG-88ER40388.

\end{document}